\documentclass[aps,pop,twocolumn,superscriptaddress]{revtex4}
\usepackage{mathrsfs}
\usepackage{amsmath}
\usepackage{bm}
\usepackage{color}
\usepackage{graphicx}
\usepackage{hyperref}
\hypersetup{pdfpagemode=UseNone,pdfstartview=FitH,pdfstartpage=1}
\newcommand{\mycolor}{black}
\begin{document}

% \leftline{Draft version, for ...}

\title{Kinetic Ballooning Mode Under Steep Gradient: High Order Eigenstates and Mode Structure Parity Transition
 }
\author{Hua-sheng Xie}
\email[]{Email: huashengxie@gmail.com} \affiliation{Fusion
Simulation Center, State Key Laboratory of Nuclear Physics and
Technology, School of Physics, Peking University, Beijing 100871,
China}
\author{Zhi-xin Lu} \email[]{Email: luzhixinpku@gmail.com} \affiliation{Max-Planck-Institut f\"ur
Plasmaphysik, Boltzmannstr. 2, 85748 Garching,
Germany}\affiliation{Fusion Simulation Center, State Key Laboratory
of Nuclear Physics and Technology, School of Physics, Peking
University, Beijing 100871, China}
%\author{Bo Li} \email[]{Email: bli@pku.edu.cn}\affiliation{Fusion Simulation Center, State Key Laboratory of Nuclear Physics and Technology, School of Physics, Peking University, Beijing 100871, China}
\author{Bo Li} \email[]{plasma@buaa.edu.cn}\affiliation{Fusion Simulation Center, State Key Laboratory of Nuclear Physics and Technology, School of Physics, Peking University, Beijing 100871, China}
\affiliation{School of Physics and Nuclear Energy Engineering, Beihang University, Beijing 100191, China}

\date{\today}

\begin{abstract}
The existence of kinetic ballooning mode (KBM) high order
(non-ground) eigenstates for tokamak plasmas with steep gradient  is
demonstrated via gyrokinetic electromagnetic eigenvalue solutions,
which reveals that eigenmode parity transition is an
intrinsic property of electromagnetic plasmas. The eigenstates with
quantum number $l=0$ for ground state and $l=1,2,3\ldots$ for
non-ground states are found to coexist and the most unstable one can
be the high order states ($l\neq0$). The conventional KBM is the
$l=0$ state. 
It is shown that the $l=1$ KBM has the same mode structure parity as the micro-tearing mode (MTM). In contrast to the MTM, the $l=1$ KBM can be driven by pressure gradient even without collisions and electron temperature gradient. The relevance between various eigenstates of KBM under steep gradient and edge plasma physics is discussed. 

\end{abstract}

\pacs{52.35.Py, 52.30.Gz, 52.35.Kt}

%\keywords{}
\maketitle
%\end{CJK*}
\section{Introduction}
Transport barrier, which forms a steep gradient profile, such as at
the edge region of high confinement mode in toroidal magnetically
confined plasmas, may not only help to make fusion economically
possible but also has fundamental interest
\cite{Hahm2002,Wagner2007}. It is known that the edge steep gradient
physics, including both electrostatic and electromagnetic
turbulence, is much more complicated and different compared with the
core weak gradient plasma. Significant progress has been made
recently to understand the edge electrostatic turbulence
(cf.\cite{Xie2017,Chang2017}). The study of electromagnetic
turbulence is more challenging {\color{black} due to the very
complicated multi-scale physics}. Due to non-zero $\beta$ (the ratio
of thermal pressure to magnetic pressure), electromagnetic
instabilities, including the ideal ballooning mode
(IBM)\cite{Connor1978} and microscale kinetic ballooning mode
(KBM)\cite{Tang1980} play central role in the EPED model
\cite{Snyder2009} to predict the width and height of the fusion
plasma pedestal. Both KBM \cite{Dickinson2012,Wan2012} and
microtearing mode (MTM) \cite{Hazeltine1975,Connor1990} are
identified to be important at the plasma edge from experiments and
gyrokinetic theory
\cite{Applegate2007,Guttenfelder2011,Zuin2011,Dickinson2012,Moradi2013,Chowdhury2016,Hatch2016},
and thus it is crucial to understand them in order to reveal the
complicated nonlinear physics at the transport barrier.
{\color{black} These mirco-instabilities also have fundamental
importantance in space physics \cite{Hameiri1991}.}

{ Another important issue is the mode structure parity
and the related structure symmetry breaking, which is relevant to toroidal momentum transport \cite{peeters2011overview,Diamond2013}, energetic particle physics \cite{Chen2016} and electron transport \cite{Applegate2007,Guttenfelder2011,Zuin2011,Doerk2011}. Specifically, the} conventional ballooning mode (BM) has even
parity of electrostatic potential perturbation $\delta\phi$ along
magnetic field line and odd parity of parallel magnetic potential
perturbation $\delta A_\parallel$ (referred as ballooning parity).
On the contrary, MTM has odd parity of $\delta\phi$ and even parity
of $\delta A_\parallel$ (referred as tearing parity). Due to its
effect on microscopic magnetic island formation and the consequent
strong electron transport, MTM has attracted significant interest
recently
\cite{Applegate2007,Guttenfelder2011,Zuin2011,Doerk2011,Dickinson2012,Moradi2013,Swamy2014,Chowdhury2016,Hatch2016}.
In contrast to the pressure gradient driven KBM, the destabilizing
mechanism for MTM has not been fully understood {\color{black} and
even can be confusing in literature}
\cite{Moradi2013,Chowdhury2016}. While tearing or microtearing
instability is affected by collision
\cite{Hazeltine1975,Connor1990}, kinetic electron effect such as
trapped particle effects\cite{Connor1990,Swamy2014}, and electron
gradient drive \cite{Hazeltine1975,Connor1990}, later experimental
observations showed other effects such as magnetic drifts and
electrostatic potential can be destabilizing \cite{Applegate2007}
and the collision dependence can be significantly different
\cite{Moradi2013} in spherical tokamak. Both KBM and MTM are high
mode number electromagnetic mode and require pressure gradient (from
the temperature/density gradient of electron/ion), which are both
important at the plasma edge. { One} unambiguous difference
between them is the mode structure parity. {Thus, one important
question is: what leads to the different mode parities?}

In this work, we {deal with the electromagnetic
microinstabilities under steep gradients using gyrokinetic model. 
We focus on the general property of the eigenmode structure of different eigenstates for specified steep gradient as to be shown in the following. 
{\color{\mycolor} The ground eigenstate with quantum number $l=0$ and the excited eigenstates with $l=1,2,\ldots$ are obtained by solving the electrmagnetic gyrokinetic equations. The most unstable eigenstate can be the high order state ($l\ne0$) for large profile gradient, e.g., large pressure gradient. The mode structure along equilibrium magnetic field line can change from even parity to odd parity when the quantum number changes from even to odd, e.g., from the ground eigenstate ($l=0$) to the first excited state ($l=1$).}
It is demonstrated that 
%both electrostatic and electromagnetic 
non-ground eigenstates can be important or even
dominant in strong gradient plasma region. 
%We also provide a possible idea to explain the unstable mechanism of MTM, which has been a critical outstanding issue, when more comprehensive model is adopted in our analyses. By considering MTM as the $l=1$ excited eigenstate, ($l=0,1,2,\cdots$ is the index for eigenstates), the previous KBM studies become meaningful to MTM excitation since the question comes to how to excite the non-ground state.} While destabilizing mechanisms vary from collisions to kinetic effects, the key ingredient of { the high order KBMs excitation and the related mode structure parity variation} is the change of the ``quantum'' potential well and the dominance of the non-ground state as the most unstable solution. It is well known that the existence condition for classical low mode number (long wavelength) tearing mode {with odd mode parity} \cite{Furth1963} is any mechanisms (collision, kinetic, etc) to break the ideal magnetohydrodynamic (MHD) frozen law. We may address here analogously { that one mechanism of the odd parity mode structure can be} any mechanism that excites high order KBM.
{\color{\mycolor}It is  also expected that this study of KBM high order eigenstates can shed light on the study of MTMs.}
This work is organized as follows. In Section \ref{sec:model}, the physics model is introduced. In Section \ref{sec:results}, the numerical results and analyses are demonstrated. In Section \ref{sec:conclusions}, we give the conclusions.

\section{Physics model}\label{sec:model}
We use collisionless gyrokinetic model \cite{Brizard2007} for our
analysis. For electrostatic and electromagnetic perturbation $\delta
\phi$ and $\delta A_\parallel$ with
$\partial_{l_\parallel}\delta\psi=i\omega\delta A_\parallel$, the
perturbed distribution function after gyrophase average is
decomposed as
\begin{equation}
    \delta F_j=\frac{q_j}{m_j}\Big(\delta\phi\frac{\partial F_{0j}}{\partial E}
    -\frac{Q}{\omega}F_{0j}J_0^2\delta\psi\Big)+J_0(\lambda_j)\delta K_j,
\end{equation}
where $\partial F_{0j}/\partial E=-m_jF_{0j}/T_j$, the subscript $j$ indicates species ($j=e,i$ for electrons and ions respectively), $T_j$ is the temperature, $m_j$ is the mass,
$QF_{0j}=(\omega\partial/\partial
E+\hat\Omega_{*j})F_{0j}=(m_j/T_j)(-\omega+\omega_{Tj})F_{0j}$,
$E=v^2/2$, $v^2=v_\parallel^2+v_\perp^2$, {\color{\mycolor}$v_\parallel$ and $v_\perp$ are velocities in parallel and perpendicular directions with respect to the equilibrium magnetic field $\bf B$, }$\omega_{Tj}=
\omega_{*j}\Big[1+\eta_j\Big(\frac{v^2}{2v_{tj}^2}-\frac{3}{2}\Big)\Big]$,
$\omega_{*j}=-ck_\theta T_j/(q_j BL_n)$, {\color{\mycolor}$k_\theta$ is the poloidal angular wavenumber, } $\mu=v_\perp^2/(2B)$,
$\Omega_j=q_jB/(m_jc)$, {\color{\mycolor}$q_j$ is the electric charge, $B$ is the equilibrium magnetic field magnitude, $\rho_j=v_{tj}/\Omega_{j}$. We have assumed
isotropic Maxwellian $F_0=n_0F_M$ with $n_0$ being the equilibrium density, }$F_M=({m_j}/{2\pi
T_j})^{3/2}e^{-m_jE/T_j}$ and $v_{tj}=\sqrt{T_j/m_j}$. The
parallel magnetic perturbation $\delta B_\parallel$ is omitted,
which is valid at low $\beta$, with $\beta\equiv{8\pi
n_0T}/{B^2}$. Linearized gyrokinetic equation for $\delta K_j$ is
given by \cite{Chen1991}
\begin{eqnarray}\label{eq:gk1d0} \nonumber
\Big(v_\parallel\frac{\partial}{\partial
l_\parallel}-\imath\omega+\imath\omega_{Dj}\Big) \delta
    K_j=
      \imath\frac{q_j}{m_j}QF_0\Big[J_0(\delta\phi-\delta\psi) \\
      +\frac{\omega_{Dj}}{\omega}J_0\delta\psi
      +\imath\frac{v_\parallel}{\omega}J_1\frac{\partial \lambda}{\partial l_\parallel}\delta\psi\Big],
\end{eqnarray}
where the imaginary number $\imath=\sqrt{-1}$, $J_0(\lambda_j)$ and $J_1(\lambda_j)$ are Bessel functions
with $\lambda_j=k_\perp v_\perp/\Omega_{j}$, and $l_\parallel$ is
the parallel coordinate represents the distance along the field
line. Quasi-neutrality equation (Poisson equation)
$\sum_jq_j\langle\delta F_j\rangle_j=0$ yields,
\begin{equation}\label{eq:qn0}
    \sum_j\frac{q_j^2}{m_j}\Big\langle \delta\phi\frac{\partial F_0}{\partial E}-
    \frac{Q}{\omega}F_0J_0^2\delta\psi +\frac{m}{q}J_0\delta K\Big\rangle_j=0,
\end{equation}
{\color{\mycolor}where $\langle\cdots \rangle=\int (\cdots)
d^3v=4\pi\int(\cdots)(B/|v_\parallel|)d\mu dE$ and the subscript $j$ following the bracket applies to species quantities inside the brackets.} Combining
quasi-neutrality equation and parallel Ampere's law, we obtain the
vorticity equation
\begin{eqnarray}\label{eq:vort0}\nonumber
    \frac{c^2B}{4\pi\omega^2}\frac{\partial}{\partial
    l_\parallel}\Big(\frac{k_\perp^2}{B}\Big)\frac{\partial\delta\psi}{\partial
    l_\parallel}=\sum_j
    q_j\Big\langle\frac{\omega_D}{\omega}J_0\delta
    K-
    \frac{\imath}{\omega}v_\parallel J_1\frac{\partial \lambda}{\partial
    l_\parallel}\delta
    K\Big\rangle_j&&\\+\sum_j\frac{q_j^2}{m_j}\Big\langle\delta\phi\frac{\partial F_0}{\partial
    E}-\frac{Q}{\omega}F_0\Big(J_0^2\delta\phi+\frac{\omega_D}{\omega}J_0^2\delta\psi\Big)\Big\rangle_j.
\end{eqnarray}
Eqs.(\ref{eq:gk1d0})-(\ref{eq:vort0}) comprise the standard
gyrokinetic system \cite{Chen1991} which we are going to solve. In
the following analysis, only passing particles are considered and
$v_\parallel$ and $v_\perp$ are assumed to be constant along field
line. Electron is treated as massless fluid by setting $\delta
K_e=0$ and $\lambda_e=0$, which means $\omega\ll k_\parallel
v_{te}$.

By using the local $s$-$\alpha$ gyrokinetic model and ballooning
representation \cite{Connor1978} with $s\equiv (r/q)(dq/dr)$ the
magnetic shear and $\theta$ the ballooning coordinate along field
line, we have $\partial_{l_\parallel}=1/(qR)\partial_\theta$, {\color{\mycolor}where $R$ is the major radius, }
$k_\perp^2=k_\theta^2[1+(s\theta-\alpha\sin\theta)^2]$,
$\omega_{Dj}=\omega_{dj\theta}(v_\perp^2/2+v_\parallel^2)/(2v_{tj}^2)$,
$\omega_{dj\theta}=\omega_{dj}[\cos\theta+\sin\theta(s\theta-\alpha\sin\theta)]$,
and $\omega_{dj}=2\epsilon_n\omega_{*j}$, $\epsilon_n=L_n/R$,
$\alpha=-2(Rq^2/B^2)dp/dr=\frac{q^2}{\epsilon_n}\beta_i[(1+\eta_i)+\tau_e(1+\eta_e)]$, {\color{\mycolor}$p$ is the total pressure, $\beta_j=8\pi n_{j}T_j/B^2$, $\tau_e=T_e/T_i$, }
$\beta_e=\tau_e\beta_i$,
$\lambda_j=k_\perp v_\perp/\Omega_j=k_\perp\rho_{j}
\frac{v_\perp}{v_{tj}}$. $\frac{\partial\lambda_j}{\partial
\theta}=k_\theta\rho_j\frac{v_\perp}{v_{tj}}\frac{(s\theta-\alpha\sin\theta)(s
-\alpha\cos\theta)}{\sqrt{1+(s\theta-\alpha\sin\theta)^2}}=v_\perp\frac{\partial
k_\perp}{\partial \theta}$, $L_n^{-1}=-d\ln n_0/dr$, $L_T^{-1}=-d\ln
T/dr$, $\eta_j=L_n/L_{T_{j}}$, $\tau=T_{e}/T_{i}$. Here $\alpha$
represent the Shafranov shift effect. With definition $\Phi =
{\color{\mycolor}\delta}\phi-{\color{\mycolor}\delta}\psi$, $\Psi = \frac{{\color{\mycolor}\delta}\psi}{\omega}$, $g_i =
K_i-F_{Mi}\Big[J_0\Phi+\omega_{Di}J_0\Psi
      +\imath\frac{v_\parallel}{q}J_1\frac{\partial
\lambda_i}{\partial\theta}\Psi\Big]$, the final eigenvalue equations
derived from Eqs.(\ref{eq:gk1d0})-(\ref{eq:vort0}) are
\begin{eqnarray}\nonumber \label{eq:gkem_e_a}
\omega
g_i=-\imath\frac{v_\parallel}{q}\frac{\partial}{\partial\theta}g_i+\omega_{Di}g_i+
      (\omega_{Di}- \\\nonumber
      \omega_{Ti})F_{Mi}\Big[J_0\Phi+\omega_{Di}J_0\Psi
      +\imath\frac{v_\parallel}{q}J_1v_\perp k'_\perp\Psi\Big]
      \\\nonumber
      -\imath F_{Mi}\frac{v_\parallel}{q}\Big[
      J_0\frac{\partial}{\partial\theta}\Phi+\omega_{Di}J_0\frac{\partial}{\partial\theta}\Psi
      +\imath\frac{v_\parallel}{q}J_1v_\perp k'_\perp\frac{\partial}{\partial\theta}\Psi\Big]\\
      -\imath F_{Mi}\frac{v_\parallel}{q}\Big[J'_0\Phi+(\omega_{Di}J_0)'\Psi
      +\imath\frac{v_\parallel}{q}v_\perp (J_1k'_\perp)'\Psi\Big],
\end{eqnarray}
\begin{eqnarray}\label{eq:qn_e_a}
\omega(\Gamma_{0i}-1)\Psi=(1+1/\tau_e)\Phi+\omega_{*i}(\Upsilon_{1i}-1)\Psi-
    \\\nonumber
    \langle
    J_0g_i\rangle-\Gamma_{0i}\Phi-\omega_{di\theta}\Delta_{1i}\Psi,
\end{eqnarray}
\begin{eqnarray}\label{eq:vort_e_a}\nonumber
    \omega\Xi=\frac{2(k_\perp^{2})'}{\beta_eq^2}\frac{\partial\Psi}{\partial
    \theta}+\frac{2k_\perp^2}{\beta_eq^2}\frac{\partial^2\Psi}{\partial
    \theta^2}+[\omega_{*i}(\Upsilon_{1i}-1)-
    \\\nonumber
    \omega_{di\theta}\Delta_{1i}]\Phi+\omega_{di\theta}\omega_{*i}[\Upsilon_{2i}+\tau_e(1+\eta_e)]\Psi\\
    -\omega_{di\theta}^2\Delta_{2i}\Psi-\frac{k_\perp^{'2}}{q^2}\Delta_{4i}\Psi-
    \Big\langle\Big[\omega_{Di}J_{0i}
    -\imath\frac{v_\parallel}{q} J_{1i}v_\perp k'_\perp\Big]
    g_i\Big\rangle,
\end{eqnarray}
where the prime is $\partial/\partial\theta$,
$\Gamma_0=e^{-b}I_0(b)$, $\Gamma_1=e^{-b}I_1(b)$, $I_{0,1}$ are the
modified Bessel functions, $b_j=k_\perp^2\rho_j^2$,
$\Delta_1=(1-\frac{b}{2})\Gamma_0+\frac{b}{2}\Gamma_1$,
$\Delta_2=\frac{(2b^2-6b+7)}{4}\Gamma_0+\frac{b(5-2b)}{4}\Gamma_1$,
$\Delta_3=k(\Gamma_0-\Gamma_1)$,  $\Delta_4=2b(\Gamma_0-\Gamma_1)$,
 $\Upsilon_1=[(1-b\eta)\Gamma_0+b\eta\Gamma_1]$,
 $\Upsilon_2=[1-\frac{b}{2}+(b-1)^2\eta]\Gamma_0+\frac{b}{2}[1+(3-2b)\eta]\Gamma_1$, and
$\Xi=(1/\tau_e)\Phi- \omega_{di\theta}\Psi-\langle J_0g_i\rangle$.
We can have an eigenvalue system $\omega AX=BX$, with
$X=[g_i,\Phi,\Psi]$. Using central difference discretization, we can
have both $A$ and $B$ to be sparse matrix. This approach can obtain
all the eigen solutions in the system, in contrast to other
iterative root finding approach and initial value codes. First, multiple roots are tracked simultaneously rather
than only the most unstable one. Second, a broad parameter regime
has been studied in which the transition between the ground state dominance to
excited state dominance is observed. Third, the parity of the mode structure
is analyzed to identify the ground and excited states and compared to the theoretical
formula. More details of the present approach can be found at
Ref.\cite{Xie2017a} of the MGK code.

\begin{figure}
 \centering
  \includegraphics[width=8.5cm]{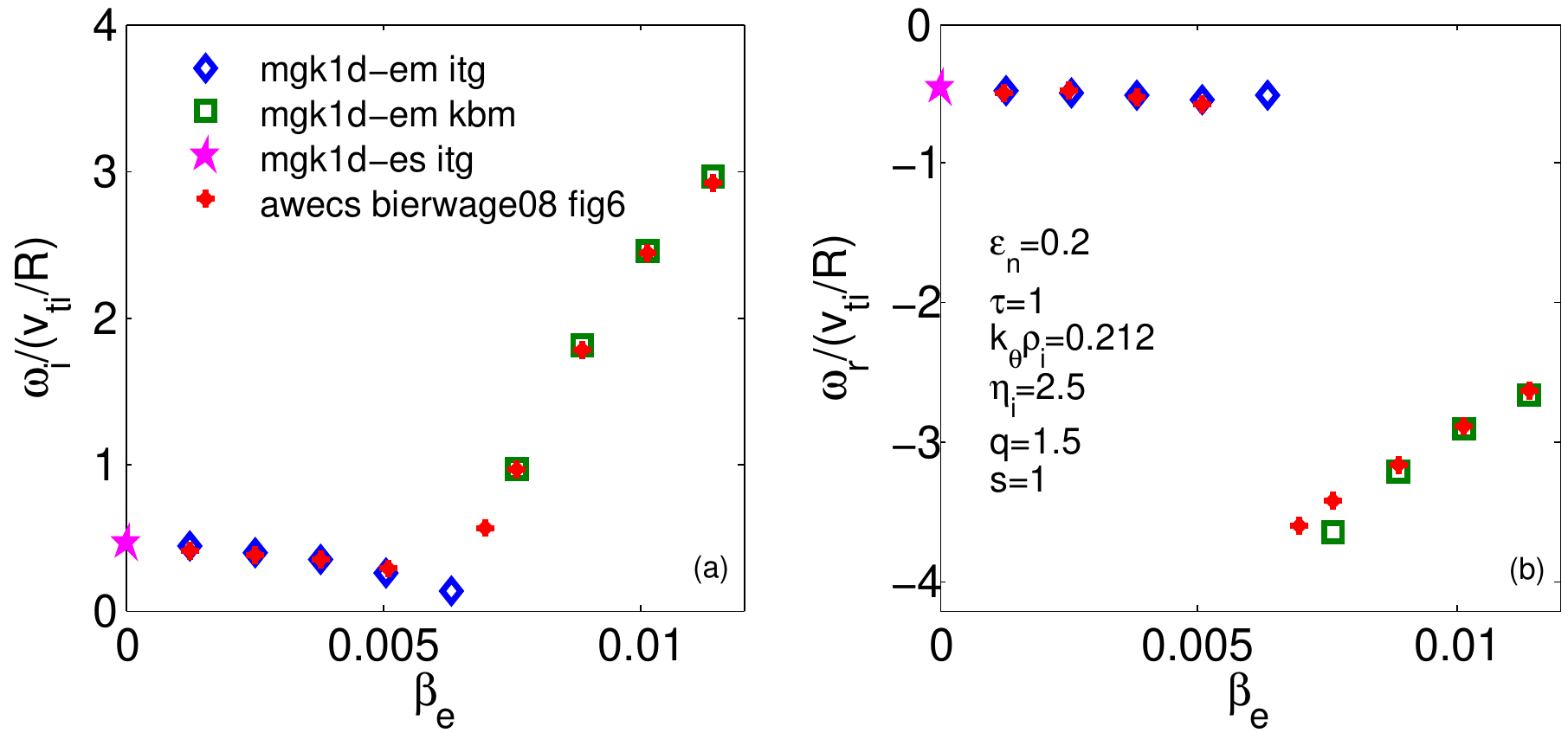}\\
  \caption{Benchmark of the eigenvalue solver MGK1d with
  particle code AWECS. {\color{\mycolor}Left and right frames show the  imaginary ($\omega_i$) and real ($\omega_r$) parts of the eigenvalue divided by $v_{ti}/R$ respectively for different values of $\beta_e$. The blue diamonds and green squares indicate the ITG and KBM branches by solving Eqs. (\ref{eq:gkem_e_a})-(\ref{eq:vort_e_a}). The magenta pentagram indicates the electrostatic ITG branch with the constraint $\Psi=0$, which is tested in \cite{Xie2017}. }}\label{fig:mgk1d_bech_awecs}
\end{figure}

\begin{figure}
 \centering
  \includegraphics[width=8.5cm]{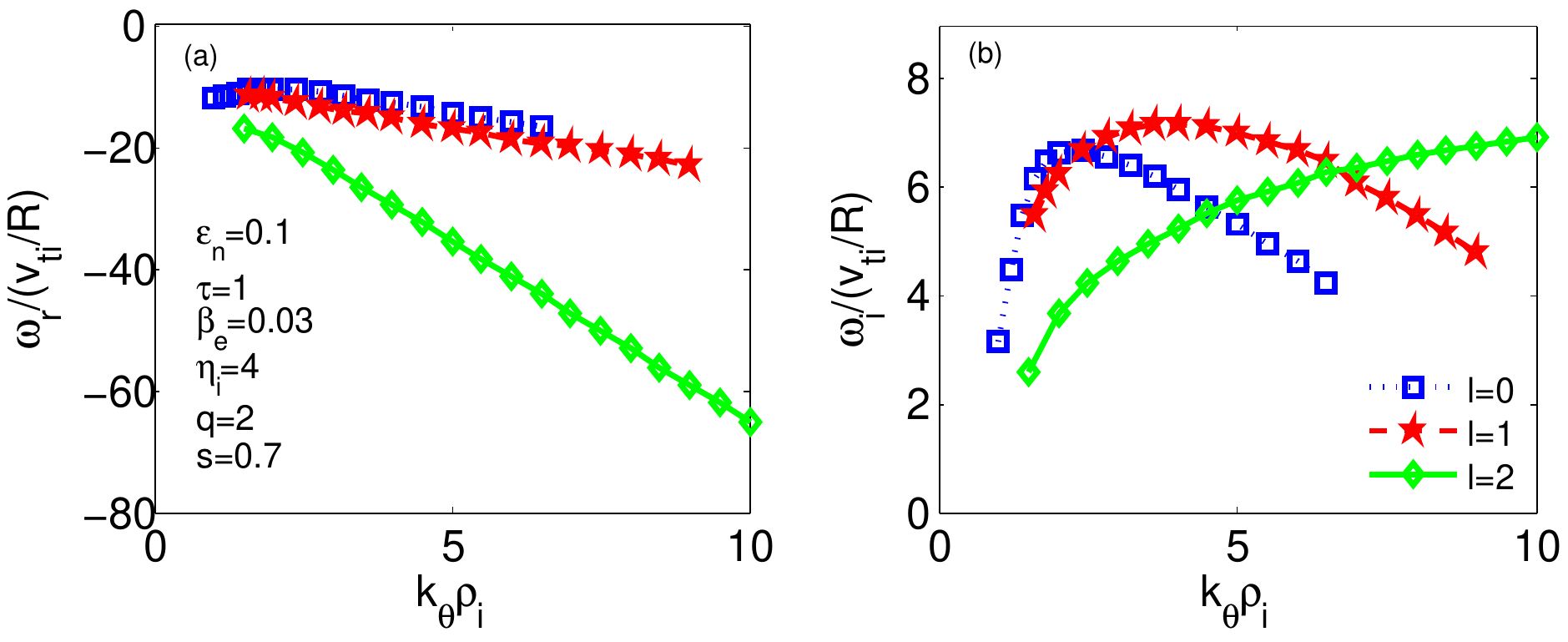}\\
  \caption{{\color{\mycolor}Left and right frames show the  imaginary ($\omega_i$) and real ($\omega_r$) parts of the eigenvalue divided by $v_{ti}/R$ respectively for different values of $k_\theta\rho_i$. Different eigenstates of KBM co-exist, labeled with $l=0,1,2$.} The $l=1$ KBM dominates for
  $3\lesssim k_\theta\rho_i \lesssim7$, and the $l=2$ KBM  dominates
  for $k_\theta\rho_i \gtrsim 7$.}\label{fig:kbm_scan_k}
\end{figure}

\begin{figure}
 \centering
  \includegraphics[width=8.5cm]{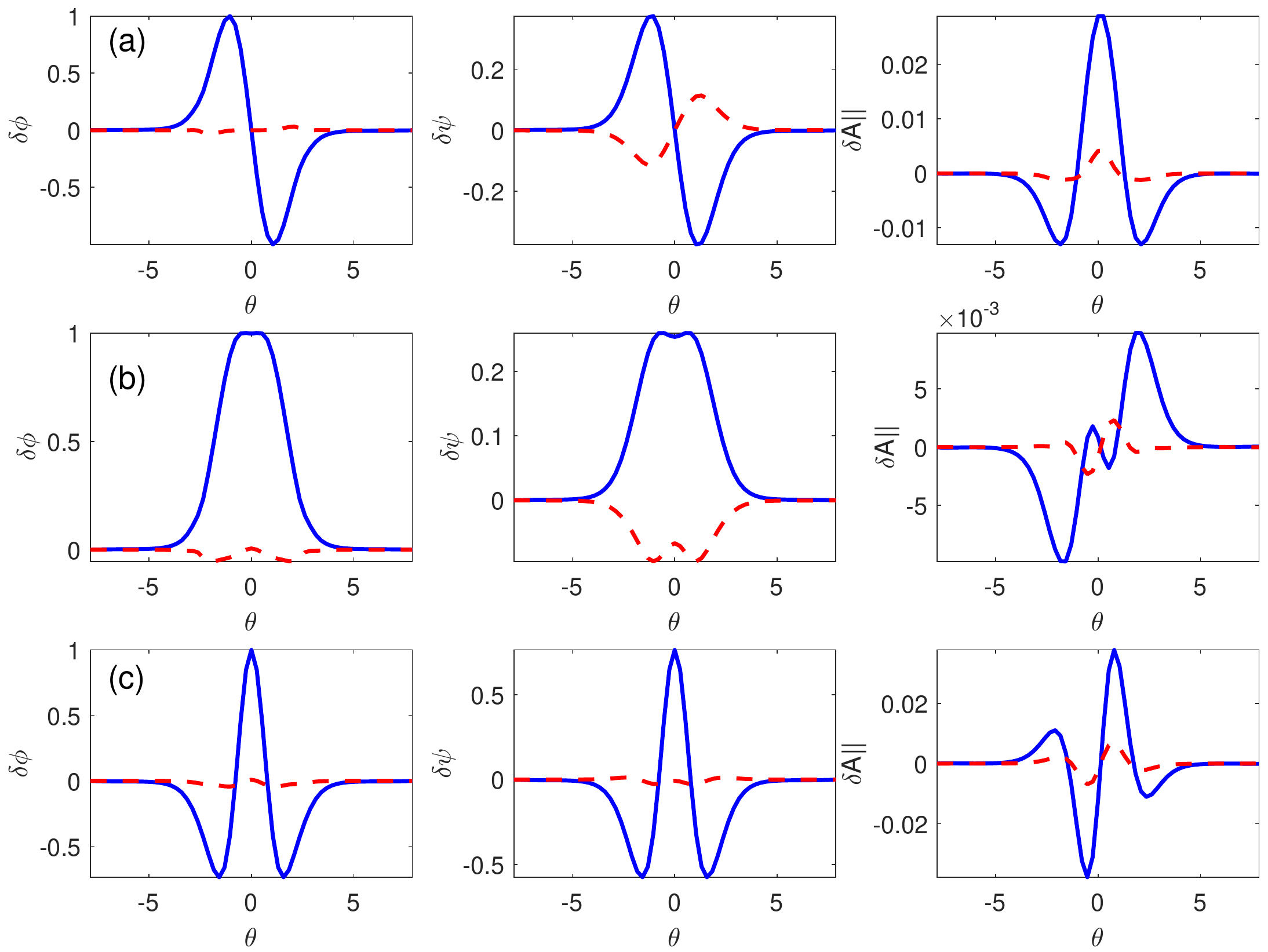}\\
    \caption{{\color{\mycolor}Mode structures of $\delta\phi$ (left), $\delta\psi$ (center) and $\delta A_\parallel$ (right) for the
  most unstable mode with $l=1$ and $\omega/(v_{ti}/R)=-17.2+7.04\imath$ (first row), the ground state mode (third most unstable) with $l=0$ and $\omega/(v_{ti}/R)=-14.5+5.31\imath$  (second row) and second most unstable mode with $l=2$ and $\omega/(v_{ti}/R)=-35.4+5.8\imath$  (third row)
  KBMs with $k_\theta\rho_i=5.0$. The blue solid line and the red dashed line indicate the real and imaginary parts of $\delta\phi$, $\delta\psi$ and $\delta A_\parallel$.}}\label{fig:kbm_phipsiA_theta}
\end{figure}

\begin{figure}
 \centering
  \includegraphics[width=8.5cm]{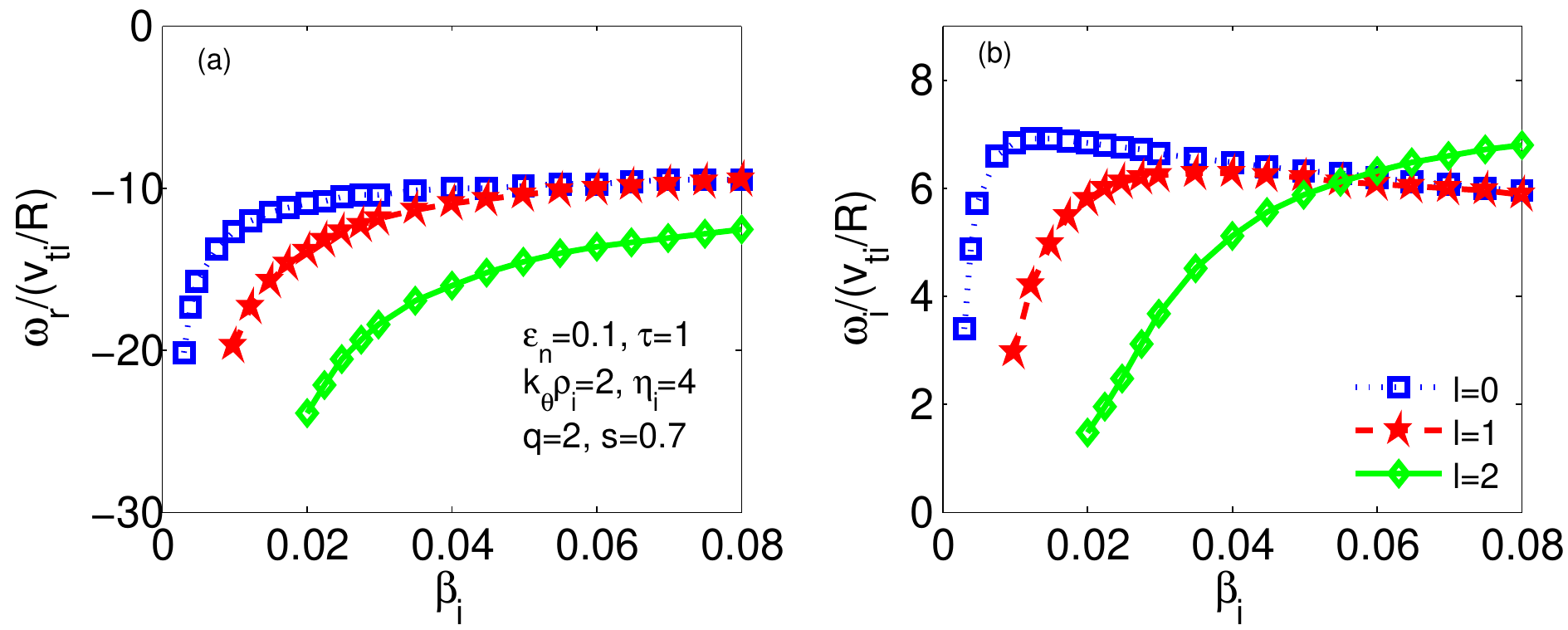}\\
  \includegraphics[width=9.3cm]{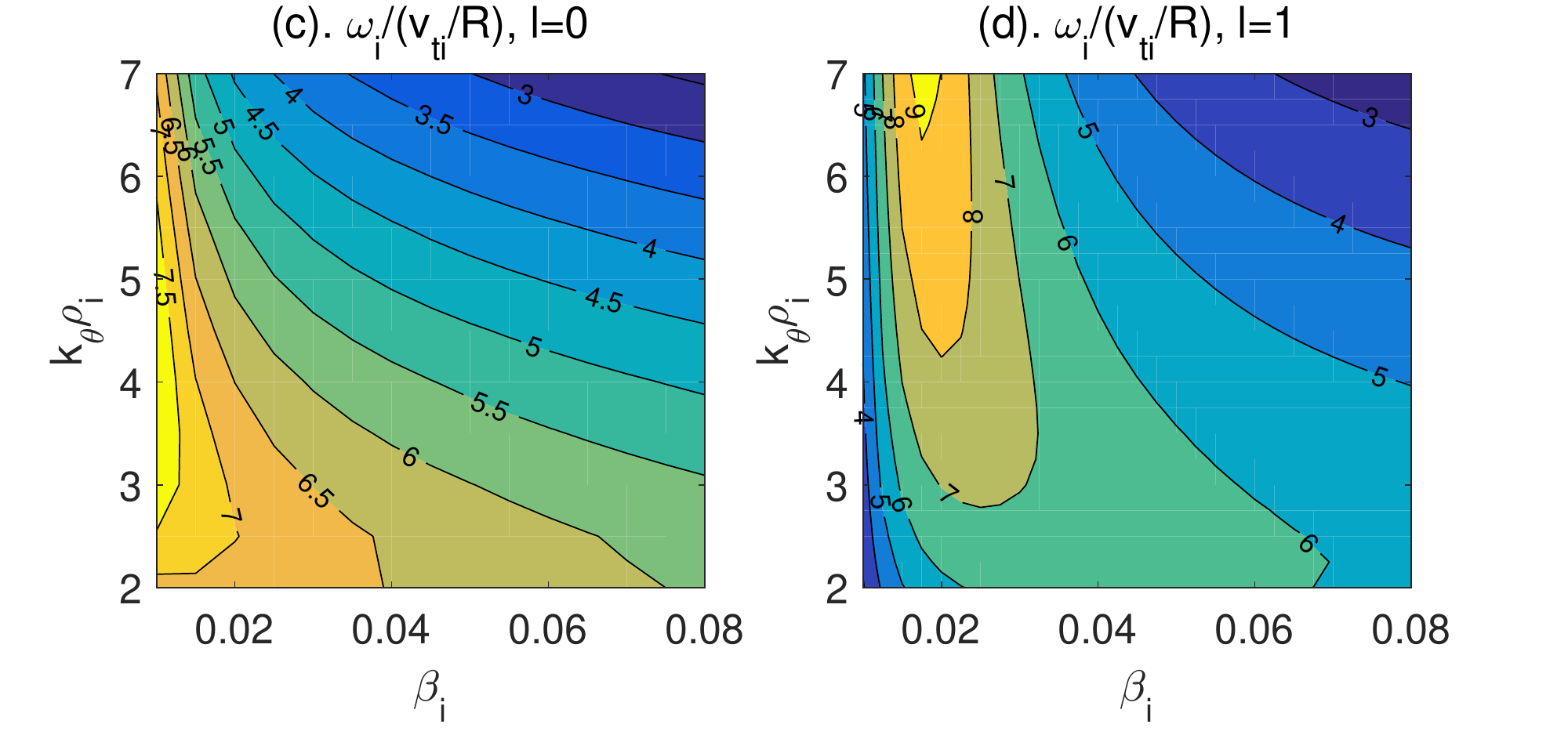}\\
  \caption{Beta scan for $l=0,1,2$ KBMs with $k_\theta\rho_i=2.0$
  (a,b) and  $(\beta_i,k_\theta\rho_i)$ scan for $l=0,1$ (c,d).
  Other parameters are the same as in Fig.\ref{fig:kbm_scan_k}. At small $\beta_i$,
  all $l=0,1,2$ growth rates increasing with $\beta_i$ increasing,
  which is characteristic of KBM.}\label{fig:kbm_scan_beta}
\end{figure}

\begin{figure}
 \centering
 \includegraphics[width=8.5cm]{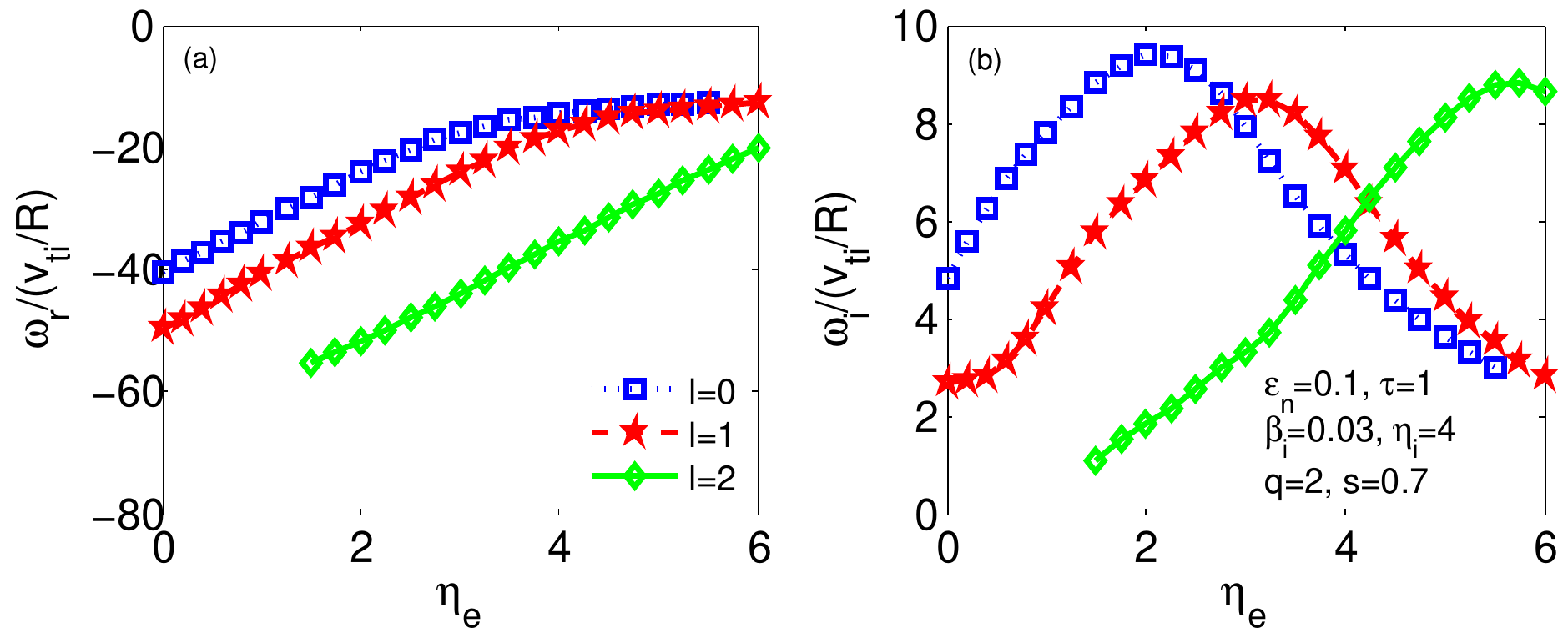}\\ 
 \includegraphics[width=9.3cm]{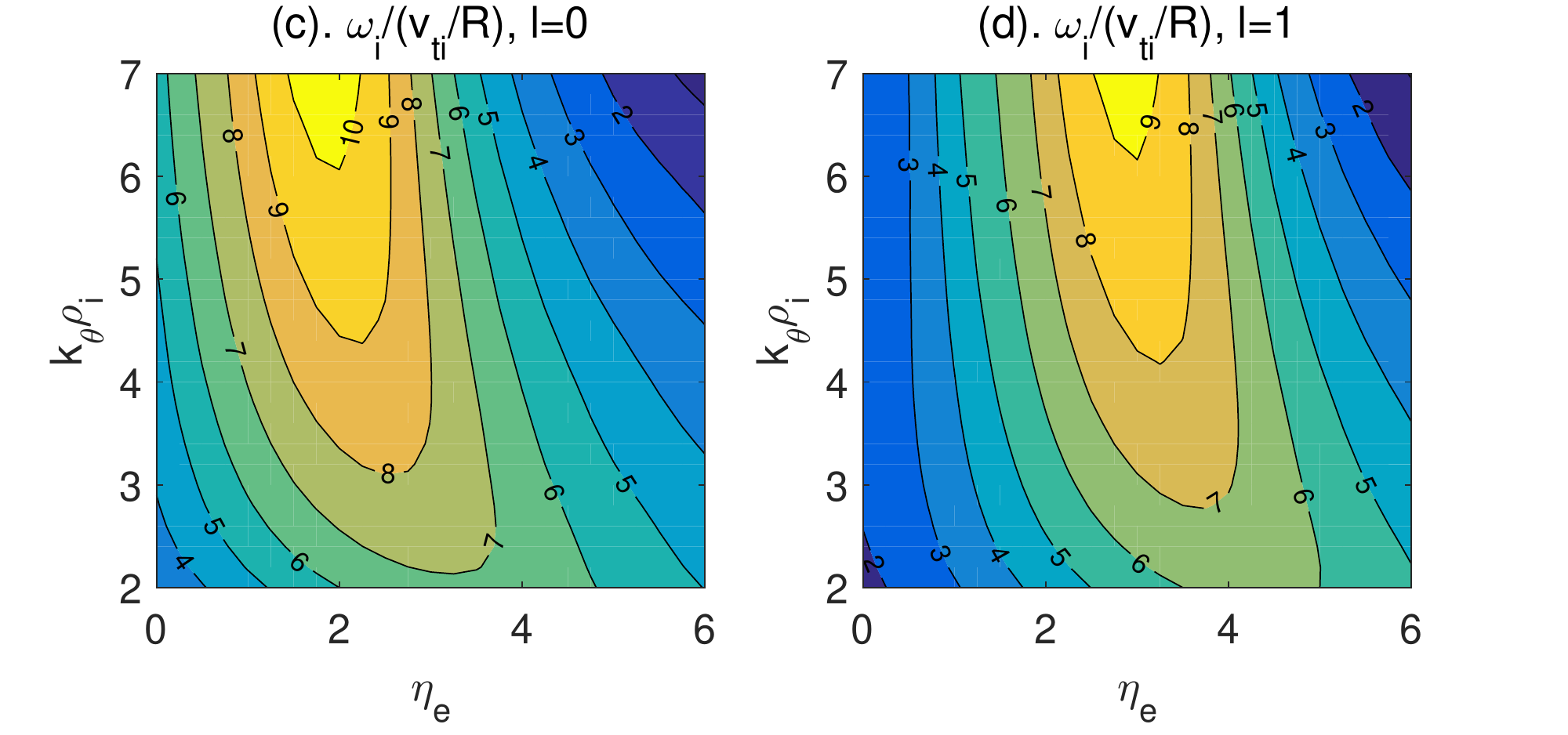}\\
  \caption{Scan of $\eta_e$ for $l=0,1,2$ (a,b) KBMs
  with $k_\theta\rho_i=5.0$ and $(\eta_e,k_\theta\rho_i)$ scan for $l=0,1$ (c,d).
  Other parameters are the same as in Fig.\ref{fig:kbm_scan_k}. For $\eta_e=0$, the $l=0,1$ KBMs are still unstable,
  which means that electron temperature gradient is not a must for $l=1$ KBM.}\label{fig:kbm_scan_etae}
\end{figure}

\begin{figure}
	\centering
	\includegraphics[width=8.5cm]{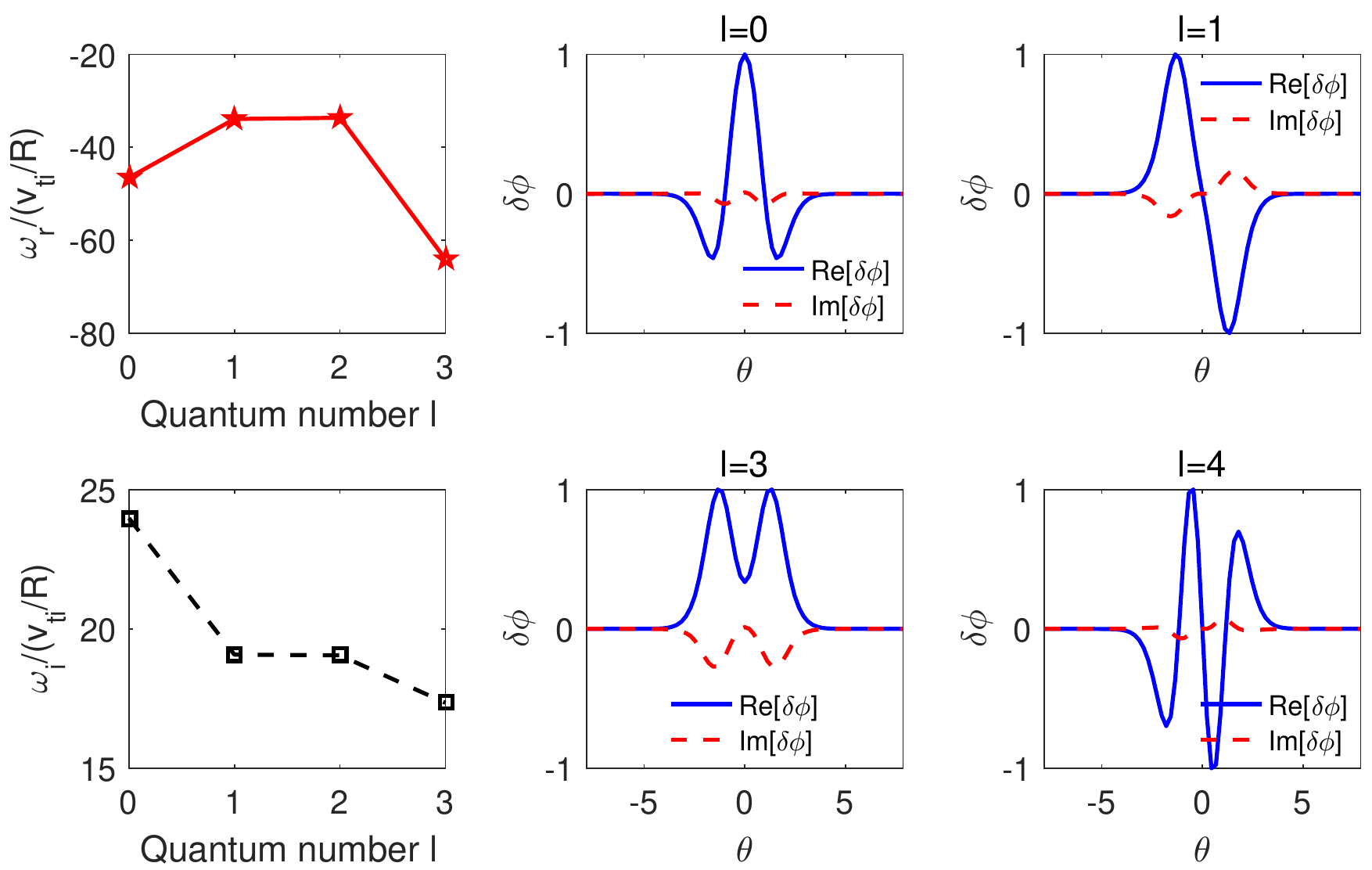}\\
	\caption{Left column of eigenvalues of $l=0,1,2,3$ for $\epsilon_n=0.05$, $\eta_i=8.5$, $k_\theta\rho_i=6.0$,
		$\beta_i=0.05$, $s=0.78$, $q=1.4$ and $\tau=1.0$. Central and right columns show the eigen functions.}\label{fig:higher}
\end{figure}

\section{Results}\label{sec:results}
We firstly benchmark our eigenvalue {\color{\mycolor}solver called MGK1d-EM which solves Eqs. (\ref{eq:gkem_e_a})-(\ref{eq:vort_e_a})} with the particle code
AWECS\cite{Bierwage2008} to repeat the conventional $l=0$ KBM.
Figure \ref{fig:mgk1d_bech_awecs} shows a $\beta_e$ scan of the
complex frequency, where good agreement is obtained. The transition
of the most unstable mode from electrostatic ITG to electromagnetic
KBM is observed clearly. At $\beta\to 0$ limit, the present
electromagnetic model can reproduce the electrostatic ITGs in
Ref.\cite{Xie2017a} as shown by the pink star in
Fig.\ref{fig:mgk1d_bech_awecs}. Hereafter, we will demonstrate our
new result for high order eigenstates of KBM under steep gradient.
Since Grad-Shafranov equilibrium with $\alpha=-2(Rq^2/B^2)dp/dr$ is not
valid for strong gradient, we will set $\alpha=0$, i.e., using
concentric circles flux surface, to study the existence of high
order KBMs. {This leads to a shift of the critical condition of eigenstate transition and the detailed discussion of finite Grad-Shafranov shift will be discussed in a dedicated work.} Note that our new high order modes are different from
the ``high order" KBM in Ref.\cite{Hirose1994}, which is still even
parity of $\delta\phi$ but has broad mode structure. And also, the
KBM in Ref.\cite{Hirose1994} is due to $\alpha$ potential well, as
studied by Ref.\cite{Hu2004}, and called $\alpha$TAE. Thus, this is
another reason we set $\alpha=0$, i.e., to remove the
$\alpha$-induced potential well.

%[Under particular parameters, positive frequency KBMs can also be
%found in the present model (not shown).]
{\color{\mycolor}
	Typical cases under steep gradient analyzed in the following are characterized by $R/L_{Ti}=40$, where  $L_{Ti}=1/(d\ln T_i/dr)$.
	Analyses based on more realistic experimental parameters are beyond the scope of this work.}
Figure \ref{fig:kbm_scan_k} shows the co-existence of ground ($l=0$)
and $l=1,2$ non-ground eigenstates for various $k_\theta\rho_i$
values. Unless explicitly indicated, other parameters in this work
are $\epsilon_n=0.1$, $\tau=1$, $\beta_i=0.03$, $\eta_i=4$,
$\eta_e=4$, $q=2$, $s=0.7$, $k_\theta\rho_i=5$. While the real
frequency linearly scales with $k_\theta\rho_i$ for all eigenstates,
the most unstable eigenstate is $l=0$ for $k_\theta\rho_i\lesssim3$
and becomes $l=1$ for $3\lesssim k_\theta\rho_i\lesssim7$ and $l=2$
for $k_\theta\rho_i\gtrsim7$. The mode structures for $l=0,1,2$ are
shown in Fig. \ref{fig:kbm_phipsiA_theta} for $k_\theta\rho_i=5$. We
identify the eigenstate number $l$ by roughly fitting the mode
structure with the foregoing Weber equation solutions. The most
unstable one is the non-ground eigenstate ($l=1$) with MTM parity.
These are electromagnetic modes since $\psi/\phi\sim0.3-0.6$ is not
small. For electromagnetic ideal MHD mode with $E_\parallel\simeq0$,
we have $\delta\psi/\delta\phi\simeq1$; and for electrostatic mode,
we have $\delta\psi/\delta\phi\simeq0$. We also notice that $\delta
A_\parallel$ should have the opposite parity as that for
$\delta\phi$ and $\delta\psi$ due to the derivative $\partial_l$ in
the relation $\partial_{l_\parallel}\delta\psi=i\omega\delta
A_\parallel$.

{\color{\mycolor}Figure \ref{fig:kbm_scan_beta} (a) shows that the absolute value of real frequency increases as $\beta_i$ increases for the $l=0,1,2$ eigen states. Figure \ref{fig:kbm_scan_beta} (b) shows the growth rate
$\omega_i$ of all eigenstate increases with $\beta_i$ at small
$\beta_i$. However, when $\beta_i$ is above a critical value,
$\omega_i$ of the ground ($l=0$) state  decreases as $\beta_i$ increases and the non-ground
eigenstate can have larger $\omega_i$ than the ground state.} Figure
\ref{fig:kbm_scan_beta} (c) and (d) show the growth rate of {the ground and excited eigenstates} in $(\beta_i,k_\theta\rho_i)$ space. {The ground and excited eigenstates} can co-exist in
general and for the parameters concerned here, as $\beta_i$
increases, the dominant instability transits from {the ground state} to {the excited state} and the
spectrum shifts from $k_\theta\rho_i\sim3$ to $k_\theta\rho_i>7$.

It is generally believed that MTM with odd mode parity is driven by electron temperature
gradient
\cite{Hazeltine1975,Applegate2007,Dickinson2012,Moradi2013,Chowdhury2016},
i.e., requiring $\eta_e\neq0$. Here, we show that  the excited {\color{\mycolor}($l\neq0$) state of KBM is
unstable under steep gradient even with $\eta_e=0$ and thus is different than the MTM}. The dependence
of eigenvalue on $\eta_e$ is shown in Fig. \ref{fig:kbm_scan_etae}
(a) and (b). The fixed ion temperature gradient $\eta_i=4$ drives
the ground (dominant) and non-ground states even $\eta_e=0$. As
$\eta_e$ increases, similar to the $\beta_i$ dependence, the most
unstable eigenstate changes from $l=0$ to $l=1$. For $\eta_e=0$, the
$l=0,1$ KBMs are still unstable, which indicates that electron
temperature gradient is not the only unstable mechanism for $l=1$
KBM. The most unstable eigenstate can shift from $l=0$
to $l=1$ depending on the strength of the drive such as $\beta_i$
and $\eta_e$. The co-existence of the even and odd parity eigenstates in
$(\eta_e,k_\theta\rho_i)$ space is shown in Fig.
\ref{fig:kbm_scan_etae} (c) and (d), where the different eigenstates have
comparable growth rate in the whole parameter regime. Also note that
the model we used is collisionless, thus {the $l\geq1$ KBMs}
can become unstable without collision. For even stronger drive, with
$\epsilon_n=0.05$, $\eta_i=8.5$, $k_\theta\rho_i=6.0$,
$\beta_i=0.05$, $s=0.78$, $q=1.4$ and $\tau=1.0$, the unstable $l=3$
KBM is also found, as shown in Fig. \ref{fig:higher}. This indicates that indeed series high order
eigenstates of KBM exist and they can be unstable easily or even be
the most unstable eigenstate under strong gradient.

In mathematical aspect, the existence of high order eigenstates is
not surprising, especially for slab geometry\cite{Pearlstein1969}.
Electrostatic high order drift modes in tokamak such as ITGs (ion
temperature gradient mode) and TEMs (trapped electron mode) have
been found to be important at strong gradient edge parameters
recently, both linearly \cite{Xie2015,Xie2016} and nonlinearly
\cite{Xie2017}. The existence of $l=1$ AITG (Alfv\'enic ITG) in slab
geometry without curvature drift has been reported in
Refs.\cite{Reynders1992} and \cite{Gao2002}. This $l=1$ mode is called MTM by Reynders \cite{Reynders1992} and called tearing parity AITG by Gao {\it et al} \cite{Gao2002}. 
The $l\geq2$ modes are
considered to be stable in their slab study. For electrostatic case
in the narrow parallel mode structure limit (strong ballooning
approximation), we have a Weber equation for the mode structure (cf.
\cite{Xie2015}), $({d^2}/{d\theta^2}+\nu-c^2\theta^2)
\delta\phi(\theta)=0$, where $\nu=\nu(\omega)$, $c=c(\omega)$ and
$\theta$ is the coordinate along a field line. The eigenvalue and
eigenfunction are $D_l\equiv\nu-(2l+1)c=0$ an
$\delta\phi=H_l(\sqrt{c}\theta)e^{-c\theta^2/2}$, where $H_l$ are
$l$-th order Hermite polynomials, with $l=0,1,2,\cdots$. The
subscript $l$ represents the quantum number and even/odd $l$
correspond to even/odd mode parity respectively. This implies that
the high order KBMs can exist, and they can be ubiquitous under
edge steep gradient parameters. {\color{\mycolor}This intuitive picture based on the Weber equation provides a way to understand the more complicated model for which a series of
	KBM high order states can exist and the even parity states can be more unstable depending on the parameters. %In principle, high order states of KBMs can be more unstable than its ground state for strong gradient edge plasmas parameters. To further support the potential well mechanism, 	we also demonstrate that the odd parity eigenstate can be driven by strong gradient insteat of  electron temperature gradient for the MTM \cite{Hazeltine1975,Moradi2013}.
}

We note that the quantitative solutions will be affected by the
electromagnetic models, e.g., with trapped particle, non-adiabatic
electron, finite $\delta B_\parallel$ and global effects. For
examples, the existence conditions and parameter dependence could be
different from the above results if additional effects are included.
However, the general conclusion can be listed as follows: (1)
Unstable higher order $l\geq1$ KBMs exist and can be dominant at
strong gradient; (2) {even and odd mode structure parities are connected to different eigenstates}; {\color{\mycolor}(3) The
pressure gradient is a sufficient condition for the destabilization of higher
order KBMs, even without collision, kinetic
electron and electron temperature gradient as demonstrated in our results.}

%\begin{figure}
% \centering
%  \includegraphics[width=8.5cm]{kbm_l3_phi.pdf}\\
%  \caption{ (a) The
%  distribution of several most unstable solutions in
%  $(\omega_r,\omega_i)$ complex plane. The corresponding mode structures
%  are shown for
%   $l=2$ (b\&c), $l=1$ (d\&e), $l=0$ (f\&g) and $l=3$ (h\&i). With $\epsilon_n=0.05$, $\eta_i=8.5$,
%  $k_\theta\rho_i=6.0$, $\beta_i=0.05$, $s=0.78$, $q=1.4$ and $\tau=1.0$. }\label{fig:kbm_l3_phi}
%\end{figure}

\section{Conclusions}\label{sec:conclusions}
In summary, we have demonstrated the existence of unstable high
order KBMs. {\color{\mycolor} It merits more effort to study the excited eigenstates of  KBMs in fusion
experiments, especially under strong gradient edge plasma
parameters. Our previous works  \cite{Xie2015} only
considered electrostatic case.} In this work, we conclude that both
electrostatic and electromagnetic non-ground eigenstates can be
important or even dominant at the strong gradient plasma edge. Since
these various dominant \cite{Xie2017} and subdominant
\cite{Hatch2013} electrostatic and electromagnetic eigenstates can
affect the nonlinear physics at the same time, the edge plasma
physics can be extremely complicated, {\color{black}which can
explain why the edge physics is very different and complicated than
that in the core plasma}. {\color{\mycolor}We expect that this picture of various eigenstates can shed light on the KBM
study for steep gradient profiles, which is relevant to the edge
electromagnetic microinstabilities and consequent nonlinear physics
including turbulent transport.} {The eigenmode parity
could be also important for other physics, such as parallel momentum
transport \cite{Diamond2013} and energetic particles driven Alfv\'en
eigenmodes (AE) \cite{Chen2016,Lauber2005}, and different parity AEs
have indeed been observed experimentally\cite{Kramer2004}. And thus,
the extension of this work to other topics such as Alfv\'en
eigenmode and momentum transport is also expected to introduce new
insight.}

% The consideration of even and
%odd parity mode can make the transport model based on quasilinear
%calculation more precise \cite{Bourdelle2007}.

{\it Acknowledgments.--} HSX would like to thank Yue-Yan Li, Jian
Bao, David Hatch and Fulvio Zonca for fruitful discussions. This
work was supported by the China Postdoctoral Science Foundation No.
2016M590008, Natural Science Foundation of China under Grant No.
11675007 and 11605186, and the ITER-China Grant No. 2013GB112006.
\\

\end{document}